\documentclass[%
 reprint,
 amsmath,amssymb,
 aps,
]{revtex4}

\usepackage{graphicx}% Include figure files
\usepackage{dcolumn}% Align table columns on decimal point
\usepackage{bm}% bold math
\begin{document}

\preprint{APS/123-QED}

\title{Absorption Process and Weak Cosmic Censorship Conjecture In Kerr Black hole}% Force line breaks with \\

\author{Shun Jiang}

\email{shunjiang@mail.bnu.edu.cn}

\affiliation{%
	Department of Physics, Beijing Normal University, Beijing, 100875, China\\}%Lines break automatically or can be forced with \\

\date{\today}% It is always \today, today,

\begin{abstract}
In this paper, we investigate the Weak Cosmic Censorship Conjecture in nearly extremal Kerr black hole by absorbing particles . In previous work, they ignore the process of black hole absorbing particles. They assume the whole particle can be absorbed and the overspinning occurs. However, it is questionable whether the whole particle can be absorbed. Therefore, we will investigate it in this paper. We consider the absorption process of a particle with finite size. During this absorption process, the black hole's parameters will change. This change will prevent the rest part of particle enter black hole. We show the part entering black hole can not overspin a nearly extremal Kerr black hole. Different from Sorce and Wald, we solve the overspinning problem for Kerr black hole without radiation effect or self-force effect. Further, we use this process to a general black hole. Under three assumptions, we show if Weak Cosmic Censorship Conjecture is valid for an extremal black hole by absorbing a particle, it will be valid for the corresponding nearly extremal black hole. In many situation, the extremal black holes satisfy Weak Cosmic Censorship Conjecture and the corresponding nearly extremal black holes don't satisfy it. Therefore, if those black holes satisfy three assumptions, these violations may be solved by absorption process.
\end{abstract}

\maketitle

%\tableofcontents

\section{Introduction}\label{section1}
Naked singularity may destroy the predictability of general relativity. To avoid it, Penrose proposed the Weak Cosmic Censorship Conjecture(WCCC) \cite{Penrose:1969pc}. Weak Cosmic Censorship Conjecture asserts that singularity should be always hidden behind black hole event horizons. A general proof is still lacking. There are several method to test WCCC. One may test WCCC by particles. In 1974, Wald gave a gedanken experiment to test WCCC \cite{Wald1974Gedanken}. By throwing a particle to extremal black hole, he showed the particle with enough angular momentum to destroy horizon can not be absorbed by black hole. However, by adding a charged particle to nearly extremal charged black hole, Hubeny found WCCC was invalid \cite{Hubeny1998Overcharging}. Similarly, T.Jacobson and T.P.Sotiriou found WCCC is invalid for nearly extremal Kerr black hole \cite{Jacobson2009Erratum}. The invalidity was resolved by considering the self-force or back-reaction \cite{PhysRevLett.105.261102}\cite{Colleoni2015Self}. For other works to check WCCC with particles, see \cite{Liu:2020cji,Zhang:2013tba,Gao:2012ca,He:2019fti,Ying:2020bch,He:2019kws,Mu:2019bim,Zeng:2019hux,Wang:2019dzl,Zeng:2019jta,Han:2019kjr,Gim:2018axz,Liang:2018wzd,Yu:2018eqq,Richartz:2011vf,Richartz:2008xm,Matsas:2007bj,Matsas:2009ww,Shaymatov:2019del,Shaymatov:2019pmn}.

One may also test WCCC by external field. The scattering of an external field to a black hole have several different features compared with those observed by particles. One of them is superradiance. When frequency and charge of external field satisfy some conditions, external field will extract conserverd quantities from black holes. In consideration of scalar, Dirac and Maxwell fields, there are various tests for WCCC \cite{Liang:2020hjz,Yang:2020czk,Goncalves:2020ccm,Hu:2020biz,Hong:2020zcf,Yang:2020iat,Gwak:2019rcz,Zeng:2019baw,Duztas:2019mxr,Hong:2019yiz,Zeng:2019aao,Gwak:2019asi,Chen:2019nsr,Chen:2018yah,Gwak:2018akg,Duztas:2016xfg,Toth:2015cda}. Recently, a new version of the gedanken experiments proposed by Sorce and Wald \cite{Sorce:2017dst}. For other works to check WCCC with new version of the gedanken experiments, see\cite{Jiang:2020xow,Jiang:2020alh,Wang:2020vpn,Jiang:2020mws,Wang:2019bml,Jiang:2019vww,Duztas:2018fbc,Jiang:2019soz,He:2019mqy}.

There is an important difference between particle and field. When we use field to test WCCC, the black hole absorbs field and the parameters can change continously. However, when we use particles, this is a discontinous process and the black hole's parameters can not change continously. In fact, one always ignore the absorption process and assume the whole particle can enter black hole. However, if we assume particle is a compact object with mass $\mu$, according to Schwarzschild black hole radius, the particle size is $>$$\mu$. One may imagine an absorption process: when part of particle enters black hole, the black hole's parameters change and the rest part may not satisfy the condition to enter the black hole. Therefore, a finite size particle may be divided into two parts. And the part enters black hole may not over-charge or over-spin the black hole. We will consider this picture in this paper. We will show particle will be divided into two parts and the part entering black hole can not over-spin black hole. Furthermore, we also consider a general black hole in this absorption process. Under three assumptions, we show when extremal black hole can not be over-charged or over-spun, the corresponding nearly extremal black hole can not be over-charged or over-spun. This conclusion is useful: most cases of overcharging or overspinning occur for nearly extremal black hole and the corresponding extremal black hole always satisfy WCCC. Therefore, it may solve these invalid problems. 

This paper is organized as follows. In Sec.~\ref{section2}, we review the overspinning problems in Kerr black hole. In Sec.~\ref{section3}, we use this absorption process to solve the overspinning problems in Kerr black hole. In Sec.~\ref{section4}, we discuss a general black hole using this absorption process. In Sec.~\ref{section5}, we give a brief conclusion.

\section{Weak Cosmic Censorship Conjecture In Kerr Black hole}\label{section2}
In this section, we briefly review the overspinning problems in Kerr black hole. Historically, Jacobson and Sotiriou \cite{Jacobson2009Erratum} firstly find a nearly extremal Kerr black hole can be over-spun if two conditions are met. One condition is the geodesic trajectory of the test particle is timelike at the horizon. The other one is $J_f>M^2_f$ where $J_f$ is the final black hole's angular momentum and $M_f$ is the final black hole's mass. The first condition is lax. It allows deeply bound orbits and one may preferably avoid those orbits. Jacobson and Sotiriou also acknowledge this issue, they give two examples show particles from afar can overspin Kerr black hole. However, they don't give the full range of overspinning orbits where deeply bound orbits are disallowed. In \cite{Colleoni_20151}, the authors firstly give the full range of overspinning orbits excluding deeply bound orbits. Therefore, we will review the overspinning condition in \cite{Colleoni_20151}. The initial configuration features a Kerr black hole of mass $M$ and angular momentum $J$ where $J=Ma<M^2$. They assume a pointlike test particle of rest mass $\mu \ll M$ is sent in on a geodesic and the particle's orbit is in the equatorial plane. Therefore the angular momentum is aligned with the spin of black hole and it seems most favourable for overspinning. Because there are two killing vector fields, we can define energy $E$ and angular momentum $L$, which are constants along the geodesic. For the geodesic approximation to make sense, they assume the particle energy $\mu E$ and angular momentum $\mu L$ satisfy $\mu E \ll M$ and $\mu L \ll J$. For a nearly extremal black hole, we have 
\begin{eqnarray}
\frac{a}{M}=1-\epsilon^2
\end{eqnarray}
where $\epsilon\ll1$. Let ${\mu}^\alpha$ denote the particle's four-velovity. In Boyer-Lindquist coodrdinates (${t,r,\theta,\phi}$), we have $\mu^\theta=0$ and two constants can be written as
\begin{eqnarray}
\dot{\mu}_t=0,\dot{\mu}_\phi=0
\end{eqnarray}
where an overdot denotes differentiation with respect to the proper time. They can be written as 
\begin{eqnarray}
E=-\xi^{\alpha}_{t}\mu_{\alpha}=-\mu_t\\
L=\xi^{\alpha}_\phi\mu_\alpha=\mu_\phi
\end{eqnarray}
where
\begin{eqnarray}
\xi^{\alpha}_t=\partial^\alpha_t\\
\xi^{\alpha}_{\phi}=\partial^\alpha_\phi
\end{eqnarray}
They are killing vector fields associated with time-translation and rotational symmetries of Kerr black hole. The pair({$E$,$L$}) parametrizes the family of geodesic and the normalization $\mu^\alpha\mu_\alpha=-1$ gives the radial equation of motion, which can be written as
\begin{eqnarray}
\dot{r}^2=B(r)[E-V_-(L,r)][E-V_+(L,r)] \label{r}
\end{eqnarray}
where $B(r)=1+a^2(r+2M)/r^3$, and
\begin{equation}
V_{\pm}(L,r)=\frac{2MaL}{Br^3}(1\pm\sqrt{1+\frac{Br^3[L^2(r-2M)+r\Delta]}{4M^2a^2L^2}}) \label{V}
\end{equation}
with $\Delta=r^2-2Mr+a^2$. As shown in \cite{Colleoni_20151}, $V_-$ is manifestly negative definite, so the factor $B(r)(E-V_-)$ in Eqs.~(\ref{r}) is manifestly postive definite. Therefore, $V_+$ plays the role of an effective potential for the radial motion. The allowed range for $E$ is $E\geq V_+(L,r)$, with an equality meaning radial turning points.\\
The effective potential $V_+(L,r)$ is important to overspinning problems. Therefore, we will pay attention to it. The stationary points of $V_+(L,r)$ outside the black hole correspond to circular orbits. They should satisfy the conditions
\begin{eqnarray}
E=V_+,\partial_rV_+=0 \label{SP}
\end{eqnarray}
Using Eqs.~(\ref{V}), we can solve $E$ and $L$ in terms of the circular-orbit radius $r=R$. The $E=E_c(R)$ and $L=L_c(R)$ can be written as
\begin{eqnarray}
E_c(R)=\frac{1-2\tilde{R}^{-1}+\tilde{a}\tilde{R}^{-\frac{3}{2}}}{\sqrt{1-3\tilde{R}^{-1}+2\tilde{a}\tilde{R}^{-\frac{3}{2}}}}\\
L_c(R)=\frac{\tilde{R}^\frac{1}{2}(1-2\tilde{a}\tilde{R}^{-\frac{3}{2}}+\tilde{a}^2\tilde{R}^{-2})}{1+3\tilde{R}^{-1}+2\tilde{a}\tilde{R}^{-\frac{3}{2}}}
\end{eqnarray}
where $\tilde{R}=R/M$,$\tilde{a}=a/M$,$\tilde{L}=L/M$.\\
The number of stationary points of $V_+$ and their location depend on $L$. When $L$ is below a critical value $L_{isco}(a)$, there are none stationary points outside black hole. There are two for $L>L_{isco}(a)$: a maximum value of $V_+$ represnting an unstable circular orbit, and a minimum represnting a stable one. The critical value $L_{isco}(a)$ represents the innermost stable circular orbit(ISCO). It is given by $L_{isco}=L_c(R_{isco})$, where the ISCO radius $R_{isco}$ can be found by solving Eqs.~(\ref{SP}) and $\partial^2_rV_+(r,L)=0$. The ISCO is also the most outer boundary of the region of unstable circular orbits. To make them explicit, we show there is a relation for event horizon's radius $R_{eh}$, photon's unstable circular orbits $R_{ph}$, particle's unstable circular orbit's radius $R_u$, ISCO radius $R_{isco}$ and stable circular orbit's radius $R_s$. It can be written as
\begin{eqnarray}
R_{eh}<R_{ph}<R_u<R_{isco}<R_s
\end{eqnarray}
As shown in \cite{Colleoni_20151}, the unstable circular orbits relate to the overspinning problems. Because particles in stable orbits can't enter black hole. The unstable circular orbits may be divided into bound($E<1$) and unbound($E>1$). The radius of the innermost bound circular orbit(IBCO) can be obtained by solving $E_c(R)=1$, giving $\tilde{R}_{ibco}=[1+(1-\tilde{a})^{1/2}]^2$. For a nearly extremal black hole, we find
\begin{eqnarray}
\tilde{R}_{eh}<\tilde{R}_{ph}<\tilde{R}_{ibco}<\tilde{R}_{isco} \nonumber\\
\tilde{R}_{eh}=1+\sqrt{2}\epsilon+O(\epsilon^2) \nonumber\\
\tilde{R}_{ph}=1+\sqrt{\frac{8}{3}}\epsilon+O(\epsilon^2) \nonumber\\
\tilde{R}_{ibco}=(1+\epsilon)^2 \nonumber \\
\tilde{R}_{isco}=1+(2\epsilon)^{2/3}+O(\epsilon^{4/3})      \label{OR}
\end{eqnarray}
With those relations, we can discuss deeply bound orbits. For paticle's angular momentum $L>L_{isco}$, there is a maximum value of $V_+$ locating at $R_{max}<R_{isco}$. We want the particle absorbed by black hole can clear such peak of the effective potential. We can achieve it by assuming the test particle's intially position $r_{out}$ satisfies %We may assume the test particle represnts a compact object with mass $\mu$. Using Schwarzschild black hole, the minimum value of the particle's diameter is $\sim$ $2\mu$. In the real world, the particle's diameter should far greater than it. Therefore, we assume it's diameter $\gg \mu$. We will see the particle relating to overspinning problems requires $\mu\sim\epsilon$. Therefore the particle's diameter is$\sim$ $\epsilon$. According to Eqs.~(\ref{OR}), the distance between event horizon and deeply bound orbits is $\sim$$\epsilon$. Therefore, it is not clear whether the particle can be initially fitted in its entirety outside the black hole. \\
\begin{eqnarray}
r_{out}>R_{isco} \label{UN}
\end{eqnarray}
As shown in \cite{Colleoni_20151}, with this condition, we can exlcude the deeply bound orbits.\\
Now, we consider the overspinning problem. For unstable circular orbits, the energy $E$ and angular momentum $L_c$ satisfy
\begin{eqnarray}
\tilde{L}_c(E)=2E+(6E^2-2)^\frac{1}{2}\epsilon \label{L_c}
\end{eqnarray}
With condition (\ref{UN}), a necessary and sufficient condition for a falling particle with specific energy $E$ and angular momentum $L$ to be captured by black hole  is
\begin{eqnarray}
\tilde{L}<\tilde{L}_c(E) \label{up}
\end{eqnarray}
The overspinning condition becomes
\begin{eqnarray}
(M+\mu E)^2<aM+\mu L
\end{eqnarray}
Using $\tilde{a}=1-\epsilon^2$ and introducing $\eta=\mu/M$, this condition becomes
\begin{eqnarray}
\epsilon^2+\eta W+\eta^2E^2<0 \label{down}
\end{eqnarray}
where
\begin{eqnarray}
W=2E-\tilde{L}
\end{eqnarray}
Giving ($E$,$\eta$,$\epsilon$), Eqs.~(\ref{down}) sets a lower bound on $\tilde{L}$ and  Eqs.~(\ref{up}) sets a upper bound on $\tilde{L}$. Our aim is to find the complete range of ({$\eta$,$E$,$\tilde{L}$}) for which the conditions (\ref{UN}) (\ref{up}) and (\ref{down}) are simultaneously satisfied.\\
As shown in \cite{Colleoni_20151}, for sufficiently small $\epsilon$ ($\epsilon<4/27$), $L\leq L_{isco}$ can not satisfy Eqs.~(\ref{UN}) or Eqs.~(\ref{down}). Therefore we need the orbits with $L>L_{isco}$. For such an orbit with given ($E$,$\eta$), $\tilde{L}$ is bounded from above via Eqs.~(\ref{up}) and below via  Eqs.~(\ref{down}):
\begin{eqnarray}
\epsilon^2+2\eta E+\eta^2E^2<\eta\tilde{L}<\eta\tilde{L}_c(E;\epsilon) 
\end{eqnarray}
The permissible interval length is $\eta\Delta_L=-\epsilon^2-\eta[2E-\tilde{L}_c(E;\epsilon)]-\eta^2E^2$. Using Eqs.~(\ref{L_c}), we find
\begin{eqnarray}
\eta\Delta_L=-\epsilon^2+\eta\epsilon\sqrt{6E^2-2}-\eta^2E^2 \label{range}
\end{eqnarray}
Overspinning conditions are satisfied if and only if we find ($E$,$\eta$,$\epsilon$) for which $\eta\Delta_L>0$.
Considering $\eta\Delta_L$ in Eqs.~(\ref{range}) as a quadratic function of $\eta$, there is a maximum value
\begin{eqnarray}
max\eta\Delta_L=\frac{\epsilon^2(E^2-1)}{2E^2} \label{max}
\end{eqnarray}
It is positive only for $E>1$. Therefore, all bound orbits($E \leq 1$ ) can not overspin.
The range of $\eta$ makes $\eta\Delta_L>0$ can be written as
\begin{eqnarray}
\epsilon\eta_-(E)<\eta<\epsilon\eta_+(E)
\end{eqnarray}
where
\begin{eqnarray}
\eta\pm=\frac{1}{\sqrt{2}E^2}[\sqrt{3E^2-1}\pm\sqrt{E^2-1}]
\end{eqnarray}
To summarize, the overspinning conditions can be written as
\begin{eqnarray}
E>1 \label{E}\\
\eta\tilde{L}_c(E,\epsilon)-\eta\Delta_L(E,\eta,\epsilon)<\eta\tilde{L}<\eta\tilde{L}_c(E,\epsilon) \label{L}\\
\epsilon\eta_-(E)<\eta<\epsilon\eta_+(E) \label{q}
\end{eqnarray}
\section{The Absorption Process in Kerr black hole}\label{section3}
In Sec.~\ref{section2}, we review the overspinning problem in Kerr black hole by absorbing a test particle. There are several assumptions in this progress. Let us focus on these assumptions.

First, one assume the particle is a pointlike particle. However, as shown in Section.~\ref{section2}, the particle's mass $\mu$ which can overspin is $\sim\epsilon$. If we consider the particle as a compact object in the real world instead of a black hole, using Schwarzschild black hole radius, we find the particle's size should $\gg \epsilon$.

Second, when one consider WCCC by absorbing a particle, one may assume black hole has several parameters($x$,$y$,$z$) and particle has several parameters($\Delta x$,$\Delta y$,$\Delta z$). They ignore the absorption process and after absorption the black hole parameters becomes ($x+\Delta x$,$y+\Delta y$,$z+\Delta z$). This is a discontinuous process and it holds when particle is a pointlike particle. However, we have shown the particle is finite size($\gg$$\epsilon$). One may imagine the absorption process: When part of particle enter event horizon, the black hole may have some changes and the condition for particle to enter it will change. Therefore, the rest part of the particle may not satisfy this condition and it can't enter the black hole. This process may solve overspinning problem without self-force or radiation effect.

As mentioned above, the particle size may not be small enough and it may lead some effect on WCCC. Therefore, let us use this absorption process to investige WCCC. Firstly, giving a nearly extremal Kerr black hole with mass $M$, angular momentum $J$. We have
\begin{eqnarray}
\frac{a}{M}=\frac{J}{M^2}=1-\epsilon^2
\end{eqnarray}
where $0<\epsilon\ll 1$. For an overspinning particle with mass $\mu$($\eta=\mu/M$), angular momentum $\mu L$ and energy $\mu E$, its parameters should satisfy Eqs.~(\ref{E}) (\ref{L}) and (\ref{q}). When we consider a absorption process, the black hole mass $M$ will change. Therefore, it is not convenient to use $\eta$ and $\tilde{L}$. We rewrite overspinning conditon Eqs.~(\ref{E}) (\ref{L}) and (\ref{q}) using $E$,$L$ and $\mu$. They can be written as
\begin{eqnarray}
E>1
\end{eqnarray}
\begin{eqnarray}
\mu L_c-\eta{\Delta_L}M^2<\mu L<\mu L_c \label{a}
\end{eqnarray}
\begin{eqnarray}
\epsilon\eta_-(E)M<\mu<\epsilon\eta_+(E)M
\end{eqnarray}
There are three parameters and if we fix $E$ and $\mu$, the allowed range of angular momentum $\mu L$ can be fixed. We want the length of angular momentum's interval takes maximum. Because in this case, it will include most kinds of overspinning particles. From Eqs.~(\ref{a}), we know this is equal to take maximum value of $\eta\Delta_L(E,\eta,\epsilon)$. Using Eqs.~(\ref{max}), we find the $\eta$ should take the value:
\begin{eqnarray}
\eta=\frac{\sqrt{6E^2-2}}{2E^2}\epsilon
\end{eqnarray}
Using $\eta=\mu/M$, we find
\begin{eqnarray}
\mu=\frac{\sqrt{6E^2-2}}{2E^2}\epsilon M \label{c}
\end{eqnarray}
Combining Eqs.~(\ref{max}) and (\ref{c}), the range of overspinning specific angular momentum $L$ can be written as
\begin{eqnarray}
L=L_c-\frac{\eta\Delta_LM^2b}{\mu}=L_c-\frac{(E^2-1)}{\sqrt{6E^2-2}}Mb\epsilon \label{LL}
\end{eqnarray}
where $0<b<1$.
For convenience, we assume the particle is composed of $N$ same slices where $N\rightarrow\infty$ and each slice has equal energy and angular momentum. We assume  the black hole can absorb $n$ slices and the rest part can't enter it. The black hole mass and angular momentum become
\begin{eqnarray}
M^{(x)}=M^{(0)}+x\mu E\\
J^{(x)}=J^{(0)}+x\mu L
\end{eqnarray}
where $x=n/N$, $M^{(x)}$ and $J^{(x)}$ represent the black hole mass and energy after absorbing $n$ slices. Using Eqs.~(\ref{c}) and (\ref{LL}), one find
\begin{eqnarray}
M^{(x)}=M(1+\frac{\sqrt{6E^2-2}}{2E}x\epsilon) \label{MM}
\end{eqnarray}
\begin{eqnarray}
J^{(x)}&&=(1-\epsilon^2)M^2 \nonumber\\
&&+[2ME+(6E^2-2)^\frac{1}{2}M\epsilon]\frac{\sqrt{6E^2-2}}{2E^2}Mx\epsilon   \nonumber\\
&&-[\frac{E^2-1}{\sqrt{6E^2-1}}M\epsilon b]\frac{\sqrt{6E^2-2}}{2E^2}Mx\epsilon \label{J}
\end{eqnarray}
Using $J^{(x)}$ and $M^{(x)}$, $\epsilon^{(x)}$ can be written as
\begin{eqnarray}
\frac{J^{(x)}}{{M^{(x)}}^2}=1-{\epsilon^{(x)}}^2 \label{ab}
\end{eqnarray}
Using Eqs.~(\ref{MM}), we find
\begin{eqnarray}
\frac{1}{{M^{(x)}}^2}=\frac{1}{M^2(1+\frac{\sqrt{6E^2-2}}{2E}x\epsilon)^2} \nonumber \\
=\frac{(1-\frac{\sqrt{6E^2-2}}{2E}x\epsilon)^2}{M^2(1-\frac{6E^2-2}{4E^2}x^2\epsilon^2)^2} \nonumber \\
=\frac{(1-\frac{\sqrt{6E^2-2}}{2E}x\epsilon)^2(1+\frac{6E^2-2}{4E^2}x^2\epsilon^2)^2}{M^2(1-(\frac{6E^2-2}{4E^2}x^2\epsilon^2)^2)^2}
\end{eqnarray}
Combing it with Eqs.~(\ref{J}), after some calculation, we find
\begin{eqnarray}
\epsilon^{(x)}=\epsilon\sqrt{1+\frac{E^2-1}{2E^2}bx-\frac{3E^2-1}{2E^2}(2x-x^2)}
\end{eqnarray}
where we omit $O(\epsilon^2)$ terms.\\
During this absorption process, ${\epsilon^{(x)}}$ can represent the state of the nearly extremal black hole: when ${\epsilon^{(x)}}$ decreases, it means the nearly extremal black hole develops towards extremal black. When ${\epsilon^{(x)}}$ increases, the black hole will develop in the opposite direction. Therefore, we want to investigate the developing direction of the black hole. Let
\begin{eqnarray}
f(x)=&&{\epsilon^{(x)}}^2 \nonumber \\
=&&(1+\frac{E^2-1}{2E^2}bx-\frac{3E^2-1}{2E^2}(2x-x^2))\epsilon^2
\end{eqnarray}
differentiating $f(x)$ with $x$, one find
\begin{eqnarray}
\frac{df(x)}{dx}=(\frac{E^2-1}{2E^2}b-\frac{3E^2-1}{2E^2}(2-2x))\epsilon^2
\end{eqnarray}
Solving $df/dx=0$, we have
\begin{eqnarray}
x_{min}=1-\frac{E^2-1}{6E^2-2}b
\end{eqnarray}
Becasue $f(x)$ is a quadratic function of $x$ and overspinning leads $f(1)<0$, we find $f(x_{min})<0$. There are two solutions ($x_1$, $x_2$) for $f(x)=0$. They satisfy $0<x_1<x_{min}<1<x_2$. The point $x_1$ menas black hole absorb $x_1$ slices of particle and become an extremal black hole. It easy to see $\epsilon(x)$ decreases monotonously in the interval $[0,x_1]$. 

In above disscusion, we show a black hole with mass $M^{(x)}$ angular momentum $J^{(x)}$ can become more extremal by absorbing slice of particle with mass $\mu/N$ energy $\mu E/N$ and angular momentum $\mu L/N$. For convenience, we introduce an angular momentum $L_d(\epsilon^{(x)})$. We need ${\epsilon^{(x)}}$ doesn't change when a black hole with mass $M^{(x)}$ angular momentum $J^{(x)}$ absorbs a slice of particle with mass $\mu/N$ energy $\mu E/N$ and angular momentum $\mu L_d/N$. It is easy to see $L>L_d(\epsilon^{(x)})$ during this absorption process. This coincide with the result $\epsilon(x)$ decreases monotonously in the interval $[0,x_1]$. This result means black hole will become more extremal by absorbing slice of this kind particle.

During this absorption process, the black hole parameters change continuously and condition for a particle to be captured by the black hole will change. Using Eqs.~(\ref{L_c}) and (\ref{up}), We find the capture condition become
\begin{eqnarray}
L<2M^{(x)}E+(6E^2-2)^\frac{1}{2}M^{(x)}\epsilon^{(x)}
\end{eqnarray}
We can label it by $L<L_u(\epsilon^{(x)})$, where
\begin{eqnarray}
L_u(\epsilon^{(x)})=2M^{(x)}E+(6E^2-2)^\frac{1}{2}M^{(x)}\epsilon^{(x)}
\end{eqnarray}
It is easy to see for the initial state $L_d<L<L_u$. Therefore, at least, black hole can absorb fisrt slice of particle. When black hole absorbs more slices, $x$ will increase and $L_d(\epsilon^{(x)})$, $L_u(\epsilon^{(x)})$ will change. The key point is black hole can absorb how many slices. It is easy to see when $L=L_u$, the slice of particle can not enter black hole. We can solve this equation and find there is a solution $x_e$ where $0<x_e<x_1$. This means slice can not enter it when black hole is still a near extremal black. This means absorption process will save WCCC. The expression of $x_e$ is very complicated and we will not give exact expression here. However, we will give another method to prove it and this method can be extended to general situations. Therefore, let us focus on it. We first consider the extremal Kerr black hole. For an extremal Kerr black hole with mass $M^e$ and angular momentum $J^e$, $L_u$ the max value of angular momentum which particle can enter it becomes
\begin{eqnarray}
L_u=2M^eE
\end{eqnarray}
We have defined an angular momentum $L_d(\epsilon^{(x)})$. We need ${\epsilon^{(x)}}$ doesn't change when the black hole with mass $M^{(x)}$ angular momentum $J^{(x)}$ absorbs a slice of particle with mass $\mu/N$ energy $\mu E/N$ and angular momentum $\mu L_d/N$.
we calculate $L_d$ for extremal black hole
\begin{eqnarray}
J^e+\frac{\mu L_d}{N}=(M^e+\frac{\mu E}{N})^2
\end{eqnarray}
To first order of $1/N$, we find
\begin{eqnarray}
L_d=2M^eE
\end{eqnarray}
We find $L_u=L_d$. This is reasonable. Because this means a particle with $L>L_d$ which can make black hole more extremal can not enter the extremal black hole. This is nothing but an extremal Kerr black hole can not be over-spun. Let us analyze this result: we start with a nearly extremal black hole where $\epsilon=\epsilon^{(0)}>0$ and $L_d(\epsilon)<L<L_u(\epsilon)$. By absorbing slices, $\epsilon^{(x)}$ decreases and $L_d$,$L_u$ change continuously. However, for extremal black hole ($\epsilon^{(x)}=0$), we find $L_d(0)=L_u(0)$. During this process we have $L>L_d$ and use particle's angular momentum $L$ is a constant, this lead to $L>L_u(0)$. Combing the initial state $L<L_u(\epsilon)$ and extremal state $L>L_u(0)$, we find there is $0<x_e<x_1$ satisfies $0<\epsilon^{(x_e)}<\epsilon^{(0)}$ and $L=L_u(\epsilon^{(x_e)})$. At this point, the near extremal black hole stops absorbing slices. Becasue of $\epsilon^{(x_e)}>0$, this means when black hole is a nearly extremal black hole, the slices can not enter it. Therefore, this particle can not overspin a black hole. This also means when taking absorption process into account, the black hole even can not become extremal by absorbing particle. At this time, the rest part of paricle can not clear the peak of potential and can not enter black hole. However, there are some slices between horzion and peak of potential. One may wonder these slices may have some effects. This will not bother us. Because we consider a particle in the real world. In this situation, its size is $\gg\epsilon$ and the distance bewteen the peak of potential and horzion is$\sim\epsilon$. Therefore, those silices between them are too small compared to the whole particle and its effect can be ignored.

By considering the absorption process, we show the whole particle can not enter it. In fact, the particle will be divided into two parts: one can enter black hole and the other can't. The black hole can not be over-spun by the part entering black hole. We solve the overspinning problem without second order effect. It is easy to see this method may be useful for a general black hole. Therefore, in next section, we will consider the general case.

\section{Weak Cosmic Censorship Conjecture In General Black Hole}\label{section4}
In this section, we consider WCCC absorption process for a general black hole. For a general black, we assume it has energy $M$ and several parameters ($A$,$B$,$C$) which can lead to naked singularity. For example, let $A$ represents angular momentum and we go back to the Kerr black hole. We consider a particle with mass $\mu$, energy $\mu E$ and one general charge of ($A$,$B$,$C$). Without loss of generality, we assume particle has charge $\mu$$A$.  We use parameter $\epsilon$ represents the state of black hole. A nearly extremal black hole means $\epsilon>0$ and extremal black hole means $\epsilon=0$. When a near extremal black hole develop to an extremal balck hole, we need $\epsilon$ decreases continously. 

For a over-charged particle($\mu$,$\mu E$,$\mu A$) to a nearly extremal black hole, when we fix ($\mu$,$E$), there will be an upper boundary $A_u$ and lower boundary $A_d$ for $A$. The upper boundary means if $A$ is too large, the particle can not enter the black hole.  We need $A_d(\epsilon)$ satisfies that $\epsilon$ doesn't change when add a slice of particle with ($\mu/N$,$\mu E/N$,$\mu A_d/N$) into black hole, where $N\rightarrow\infty$.

We use the absorption process we develop in Sec.~\ref{section3} to consider WCCC. The upper boundary condition and lower boundary condition can be written as functions of $\epsilon$, we label them by $A_u(\epsilon)$ and $A_d(\epsilon)$. If we assume $A>A_d$ in this absorption process, then we can show a WCCC relation between extremal black hole and nearly extremal black hole. $A>A_d$ is our first assumption. Let us first consider what this condition means. For $A>A_d$, when each slice of over-charge particle enters black hole, the nearly extremal black hole become more extremal. This seems reasonable for an over-charged particle but we don't prove it and make it as an assumption. With this assumption we can prove that if an extremal black hole can not be over-charged, an near extremal black hole can also not be over-charged in this absorption process. The proof is same as Kerr black hole. If the initial near extremal black hole with $\epsilon$(initial)$>0$ can be over-charged by absorbing a whole particle($\mu$,$\mu E$,$\mu A$), $A$ should satisfy $A<A_u$($\epsilon$(intial)) and we assume $A>A_d$ during this process. It is easy to see when $\epsilon=0$, the condition for an extremal black hole can not be over-charged becomes $A_d(0)\ge A_u(0)$. Using $A>A_d$, $A_d(0)\ge A_u(0)$ and $A_d$($\epsilon$(intial))$<A<A_u$($\epsilon$(intial)), we find $A_u$(0)$<A<A_u$($\epsilon$(intial)). Using $A$ is a constant, $\epsilon$ decrease in this process and $A_u(\epsilon)$ is a continous funtion of $\epsilon$, we find there is $\epsilon(c)>0$ satisfies $A=A_u(\epsilon(c))$, where $0<\epsilon(c)<\epsilon$(intial). This means rest part of particle can not enter black hole. At this time, the black hole is still nearly extremal. Therefore, a nearly extremal black hole can not be over-charged when the extremal black hole can not be over-charged. In fact, it can even not become extremal. 

A similar confusion should be cleared up. We assume the particle's size is much greater than the distance between the peak of potential and horizon. Therefore, the part between them will not have influence on the result. This is reasonable. For spherical black hole, this distance of many balck holes are $0$. For Kerr black hole, we see this distance is $\sim\epsilon$ and the overspinning particle mass is$\sim\epsilon$. We consider the particle as a compact object in the real world instead of a black hole. Using Schwarzschild black hole radius, we find the particle's size should $\gg \epsilon$. Therefore, it also will not affect the result. We assume similar situations occur for other rotating black holes. This is the second assumption.

In this proof process, we only consider a particle with one general charge. This is the third assumption. If the particle has two kind charge, we can't use this general proof. But it doesn't mean the absorption process can't solve this kind problems. We can also use the method developed in Sec.~\ref{section3} to investigate WCCC.

If three assumptions holds, this conclusion is useful. First, if extremal black hole can not be over-charged by absorbing particle, then we don't need to check nearly extremal black hole. Because it also can not be over-charged. Second, we do not have any limit on interval between upper boundary $A_u$ and lower boundary $A_d$ mentioned in the second paragraph in this section. This interval is important for radiation effect and self-force. Because those are second order effect of $\epsilon$ in Kerr black hole. However, this absorption process does not depend on it.

\section{conclusion}\label{section5}
In this paper, we consider the absorption process. We use it to check WCCC and solving overspinning problems in Kerr black hole. By simulating the process of black hole absorbing a particle, we show the whole particle can not enter it. The particle will be divided into two parts: one can enter black hole and the other can't. The black hole can not be over-spun by absorbing the part entering black hole. According to it, we find an interesting fact that an nearly extremal black hole can not become extremal in this absorption process. We also use this method to a general black hole. Under three assumptions, we find a relation bewteen a general extremal black hole and nearly extremal black hole: if WCCC is valid for extremal black hole by absorbing particle, it will also be valid for near extremal black hole by absorbing particle. This gives a general relation between extremal black hole and nearly extremal black hole in Weak Cosmic Censorship Conjecture. We also show our method is different from the method considering radiation effect and self-force effect. 
\begin{acknowledgments}
This research was supported by NSFC Grants No. 11775022 and 11873044.
\end{acknowledgments}

\end{document}